\documentclass[prl,twocolumn]{revtex4}
\usepackage{graphicx}
\usepackage{color}
\begin{document}
\title{Size and shape dependence of
  finite volume Kirkwood-Buff integrals}
\author{Peter Kr\"{u}ger}
 \affiliation{Graduate School of Engineering and
 Molecular Chirality Research Center, Chiba University,
1-33 Yayoi-cho, Inage,
 Chiba 263-8522, Japan}  
{\small e-mail: {\tt pkruger@chiba-u.jp}}
\author{Thijs J. H. Vlugt}
\affiliation{
  Engineering Thermodynamics, Process \& Energy Department, Delft University of Technology,
  Leeghwaterstraat 39, 2628CB Delft, The Netherlands}
\date{\today}
\begin{abstract}
  Analytic relations are derived for finite volume integrals over the
  pair correlation function of a fluid, so-called Kirkwood-Buff
  integrals. Closed form expressions are obtained for cubes and cuboids,
  the system shapes commonly employed in molecular simulations.
  When finite volume Kirkwood-Buff integrals
  are expanded over inverse system size,
  the leading term depends on shape only through the surface area
  to volume ratio. This conjecture is proved for arbitrary shapes
  and a general expression for the leading term is derived.
  From this, a new extrapolation to the infinite volume limit
  is proposed, which converges much faster with system size
  than previous approximations and thus significantly simplifies
  the numerical computations.
\end{abstract}

\maketitle
Density fluctuations play a prominent role in statistical physics of
fluids as they are directly related to various
thermodynamic quantities such as the isothermal compressibility
and partial molecular volumes~\cite{bennaim}.
Density fluctuations can be expressed as volume integrals over
the pair correlation function (PCF), so-called Kirkwood-Buff
integrals (KBIs), which provide a direct link between structure and
thermodynamics~\cite{kb}.
Originally, the Kirkwood-Buff theory was formulated in the
infinite volume limit,
where the KBIs reduce to simple radial integrals over the
PCF~\cite{kb,bennaim}.
In practice, however, the PCF is known only up to some finite radius~$L$,
and the infinite radial integral must be truncated.
The convergence of these ``running'' KBIs with the cut-off radius~$L$
is very bad, which is a major problem of KBI
theory~\cite{bennaim,gee11,ganguly13,chiba16}.
The poor convergence is mainly due to the fact that the
oscillations of the PCF, which reflect the short-range order,
become strongly amplified by integration over volume.
Although this problem has been investigated with a variety of
approaches~\cite{nichols09,ganguly13,cortes16},
a clear understanding of the large oscillations
of the running KBI, and a simple scheme for improving the
convergence are still missing.
Recently, we have generalized KBI theory to finite
sub-volumes~\cite{kruger13,noura1}.
As expected physically, finite volume KBIs approach the
thermodynamic limit very smoothly and scale as the inverse
system size.
We have shown that for (hyper-) spheres,
finite volume KBIs can be transformed exactly to simple radial
integrals over the PCF times an analytic weight function
and have thereby extended the simplicity of the original infinite volume KBIs
to finite volumes~\cite{kruger13}.
Non-spherical volumes are also important in practice,
especially cubes and cuboids, which are the standard shapes
used in molecular simulations with periodic boundary conditions.
The shape dependence of finite volume KBI
has recently been studied numerically in the small system
approach~\cite{schnell11,strom17} and with finite volume KBI
theory~\cite{noura2}.
It was speculated on thermodynamic grounds and confirmed
numerically for a number of simple shapes,
that in the large volume limit,
finite volume KBIs depend on shape only through $A/V$,
the surface area over volume ratio~\cite{strom17,noura2}.
To date, all findings for non-spherical volumes have
been obtained with numerical simulations.
Analytic results are
only known for spheres~\cite{kruger13}.

Here, we derive analytic expressions for finite volume KBIs
for cubes and cuboids.
We rigorously prove the recent conjecture~\cite{strom17}
that the leading term in a $A/V$ expansion of finite volume KBIs
is shape independent and we derive a general expression for this surface term.
From these results, we develop a simple
extrapolation to the infinite volume limit,
which converges much faster than all previous approaches,
and thus effectively solves the long-standing convergence problem of KBIs.

We consider a fluid with PCF $h(r_{12})=g(r_{12})-1$,
where $r_{12}=|{\bf r}_1-{\bf r}_2|$, and $g(r)$ is the
radial distribution function.
The integral
\begin{equation}\label{GV}
G(V) = \frac{1}{V}\int_V d{\bf r}_1 \int_Vd{\bf r}_2
h(r_{12}) \,,
\end{equation}
yields the average particle density fluctuations in
the volume~$V$~\cite{kb},
\begin{equation}\label{fluct}
  G(V)=\frac{V}{\langle N\rangle}\left(
  \frac{\langle N^2\rangle - \langle N\rangle^2}{\langle N\rangle} -1
  \right)\;. 
\end{equation}
Here, $N$ is the number of particles, and
$\langle \dots\rangle$ denotes a grand-canonical ensemble average
restricted to the volume~$V$.
For simplicity, Eqs.~(\ref{GV},\ref{fluct}) are written for a
monocomponent fluid, but the theory can readily be extended to multicomponent
systems~\cite{noura1}.
Also note that all results obtained here are not limited to the PCF,
but $h(r)$ may be replaced by any radial two-point function that goes to zero
faster than $r^{-3}$ for $r\rightarrow\infty$.
An example is the product of the radial distribution function
and the pair potential,
in which case Eq.~(\ref{GV}) represents the interaction energy inside~$V$.
In the thermodynamic limit, $V\rightarrow\infty$,
the six-dimensional integral in Eq.~(\ref{GV})
simplifies to the simple radial integral
\begin{equation}\label{Ginfty}
G^\infty = \lim_{V\rightarrow\infty}G(V)=\int_0^\infty  h(r) 4\pi r^2dr \;,
\end{equation}
which is known as the KBI~\cite{kb}.
Interestingly, the reduction of Eq.~(\ref{GV}) to a single radial
integral can formally be achieved also for finite volumes~$V$ of
any shape. Indeed, we can rewrite Eq.~(\ref{GV}) as
\begin{equation}\label{Gw}
G(V) = \int_0^\infty h(r) w(r) dr \;,
\end{equation}
where
\begin{equation}\label{wr}
w(r) = \frac{1}{V}\int_Vd{\bf r}_1\int_Vd{\bf r}_2 \delta(|{\bf r}_1 - {\bf r}_2| - r)
\end{equation}
is a purely geometrical weight function.
It is obvious from this definition, that $w(r)$ is continuous and
of finite support, i.e.\ we have $w(r)=0$ for $r>r_{\rm max}$ where
$r_{\rm max}$ is the largest distance between two points inside~$V$.
We define 
\begin{equation}\label{tau}
\tau({\bf r}) = \int_Vd{\bf r}_1\int_Vd{\bf r}_2 \delta({\bf r}_1 - {\bf r}_2 - {\bf r}) \;,
\end{equation}
which is the overlap between the volume $V$ and
the same volume shifted by ${\bf r}$. We have
\begin{equation}\label{wt}
  w(r) = r^2 T(r)/V \;,\quad T(r)\equiv \int d\Omega\; \tau(r,\Omega) \;.
\end{equation}
For a sphere of diameter~$D$,
$w(r)$ is given by~\cite{giorgini,kruger13}
\begin{equation}\label{sphere}
w(r)
= 4\pi r^2\left(1 - \frac{3}{2}x + \frac{1}{2} x^3 \right) \;, \quad x=r/D\;.
\end{equation}
For non-spherical shapes, however,
analytic results for $w(r)$ have not been reported.

\begin{figure}
\begin{center}
\includegraphics[width=\columnwidth]{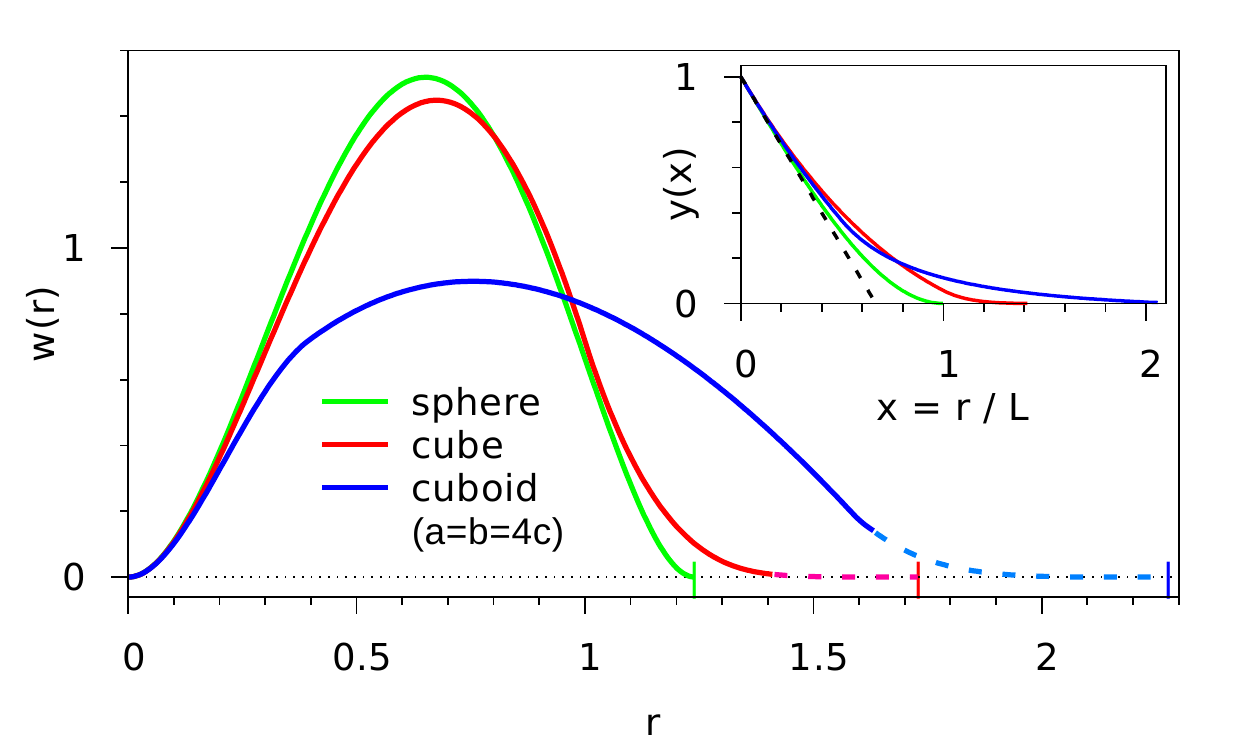}
\end{center}
\caption{
  The function $w(r)$ for a 
  sphere, a cube and a cuboid with $a=b=4c$, all with $V=1$.
  $w(r)=0$ for $r>r_{\rm max}$ (indicated as vertical bars).
  Colored broken lines are $(r_{\rm max}-r)^5$ interpolations
  in the regions where $w(r)$ is not known analytically.
  The inset shows $y(r)=w(r)/(4\pi r^2)$ as a function of $x=r/L$,
  where $L\equiv 6V/A$. For the three shapes,
  $L$ equals the diameter ($D$), side length ($a$) and harmonic mean of
  the side lengths ($3/(1/a+1/b+1/c)$) respectively.
  The black broken line is $1-3x/2$.
}\label{figwr}
\end{figure}

Here, we derive analytic expressions of the function $w(r)$ for
a general cuboid with side lengths $a \ge b \ge c$, by integrating
the overlap integral $\tau({\bf r})$ and $T(r)$ in Eqs~(\ref{tau},\ref{wt}).
The details of the calculation are given in the Supplemental
Material~\cite{sm}.
The expressions of $T(r)$
are listed in Table~\ref{tabT} and Eqs~(\ref{P},\ref{Q}).
\begin{table}  
  \begin{tabular}{l|l}
 $r$-domain & $T(r)$ \\ \hline
$0<r<c$ &  $P(r)$\\
$c<r<b$ & $P(r)-Q_{abc}(r)$\\
      $b<r<{\rm min}\{a,\sqrt{b^2+c^2}\}$
    & $P(r)-Q_{abc}(r)-Q_{cab}(r)$\\      
$a<r<\sqrt{b^2+c^2}$ & $P(r)-Q_{abc}(r)-Q_{cab}(r)-Q_{bca}(r)$\\
$\sqrt{a^2+b^2+c^2}<r$ & 0 \\
\end{tabular}
\caption{The function $T(r)$ for cuboids, in terms of $P(r)$, $Q_{abc}(r)$
  given in Eqs~(\protect\ref{P},\protect\ref{Q}). $Q_{cab}(r)$ and $Q_{bca}(r)$
  are obtained from $Q_{abc}(r)$ by permutation of $abc$.
    $a\ge b\ge c$ is assumed. Depending on the values of $a,b,c$,
  some of the $r$-domains may be empty.
} \label{tabT}.
\end{table}
\begin{equation}\label{P}
P(r)
= {4\pi}abc
- {2\pi}(ab+ac+bc)r
+ \frac{8}{3}(a+b+c)r^2
- r^3 \;.
\end{equation}
\[
Q_{abc}(r)  = 4\pi abc -{2\pi}ab r + \frac{8}{3}cr^2 -r^3 
\]
\vspace*{-2em}
\[
-\left(  {4\arccos(c/r)}(a+b)c + 2c^2\right)r 
+ \left(c^4/3-{2\pi}abc^2\right) \frac{1}{r}
\]
\vspace*{-2em}
\begin{equation}\label{Q}
+\frac{4}{3}(a+b)(c^2+2r^2) \sqrt{1-\frac{c^2}{r^2}} \;. 
\end{equation}
They provide, together with Eq.~(\ref{wt}) and $V=abc$,
the exact analytic expressions of $w(r)$ for cuboids
in the range $0<r<\sqrt{b^2+c^2}$.
No simple expression was found 
for the range $\sqrt{b^2+c^2}< r < \sqrt{a^2+b^2+c^2}=r_{\rm max}$.
However, we shall see below that in this range,
$w(r)$ is negligible except for large aspect ratios.
For the important case of a cube ($a$=$b$=$c$), we find
\begin{equation}\label{cube1}
  w(r) = {4\pi} r^2
  \left(1  - \frac{3}{2} x  + \frac{2}{\pi} x^2 - \frac{1}{4\pi} x^3 \right)
\end{equation}
for $x\equiv r/a <1$ and
\[
w(r) = r^2\left( -8\pi + 6x + 2x^3 +(6\pi -1)/x \right.
\]\vspace{-2em}
\[
+  \left. 24 x \arccos(1/x) - 8(2x^2+1)\sqrt{1-1/x^2}  \right)
\]
for $1<x <\sqrt{2}$.
The functions $w(r)$ for a sphere (Eq.~\ref{sphere}),
a cube, and a cuboid with $a=b=4c$,
all of unit volume, are compared in Fig.~\ref{figwr}.
From the inset of this figure and Eqs~(\ref{sphere},\ref{P},\ref{cube1})
it is clear that the leading behavior is $4\pi r^2(1-3x/2)$
in all three cases.
Here, $x\equiv r/L$, where $L=6V/A$ and $A$ is the
surface area.
Below, we prove that this leading behaviour holds for any shape.
The sphere has the smallest maximum distance
$r_{\rm max}$ among all possible shapes of given volume.
As seen in Fig.~(\ref{figwr})
the cube has a larger $r$-range than the sphere,
but the functions $w(r)$ are rather similar.
In the region $\sqrt{2}<r/a<\sqrt{3}$
(red dashed line) where the analytic
expression is unknown, 
$w(r)$ is negligible.
For the unit volume cuboid with $a=b=4c$ (blue line),
$w(r)$  has a much more extended $r$-range.
In the unknown region $\sqrt{b^2+c^2}<r<\sqrt{a^2+b^2+c^2}$,
$w(r)$ is small, but not obviously negligible.
However, for $r\rightarrow r_{\rm max}$,
$w(r)$ approaches zero as $\sim \Delta^5$ where
$\Delta =r_{\rm max}-r$.
This is because the overlap~$\tau({\bf r})$
is reduced to the box corners
with volume $\sim\Delta^3$ and the solid angle integration in
Eq.~(\ref{wt}) yields a factor $\sim\Delta^2$.
The $\Delta^5$ scaling can be used as a simple and good interpolation
of the unknown region,
as indicated in Fig.~\ref{figwr} with the blue dashed line.

Equations~(\ref{sphere},\ref{cube1}) and the inset of Fig.~\ref{figwr}
suggest that for small $r$, $w(r)$ only depends
on the surface area over volume ratio $A/V$.
Here we prove that this holds for any shape.
As a direct corollary, we then prove the conjecture~\cite{strom17} that
the leading term of $G(V)$ in a large volume expansion,
depends only on $A/V$
and we derive a general expression for this term.

{\it Theorem.}\/
For a closed volume~$V$ with a piecewise smooth
surface of area $A$, the series expansion of $w(r)$
around $r=0$ is given by
\begin{equation}\label{wtheorem}
w(r) = 4\pi r^2 \left( 1 - \frac{A}{4V}\; r  + {\cal O}(r^2) \right) \;, 
\end{equation}
independently of the shape of the volume. 

\begin{figure}
\begin{center}
\includegraphics[bb = 20 150 750 550,clip=true,width=0.8\columnwidth]{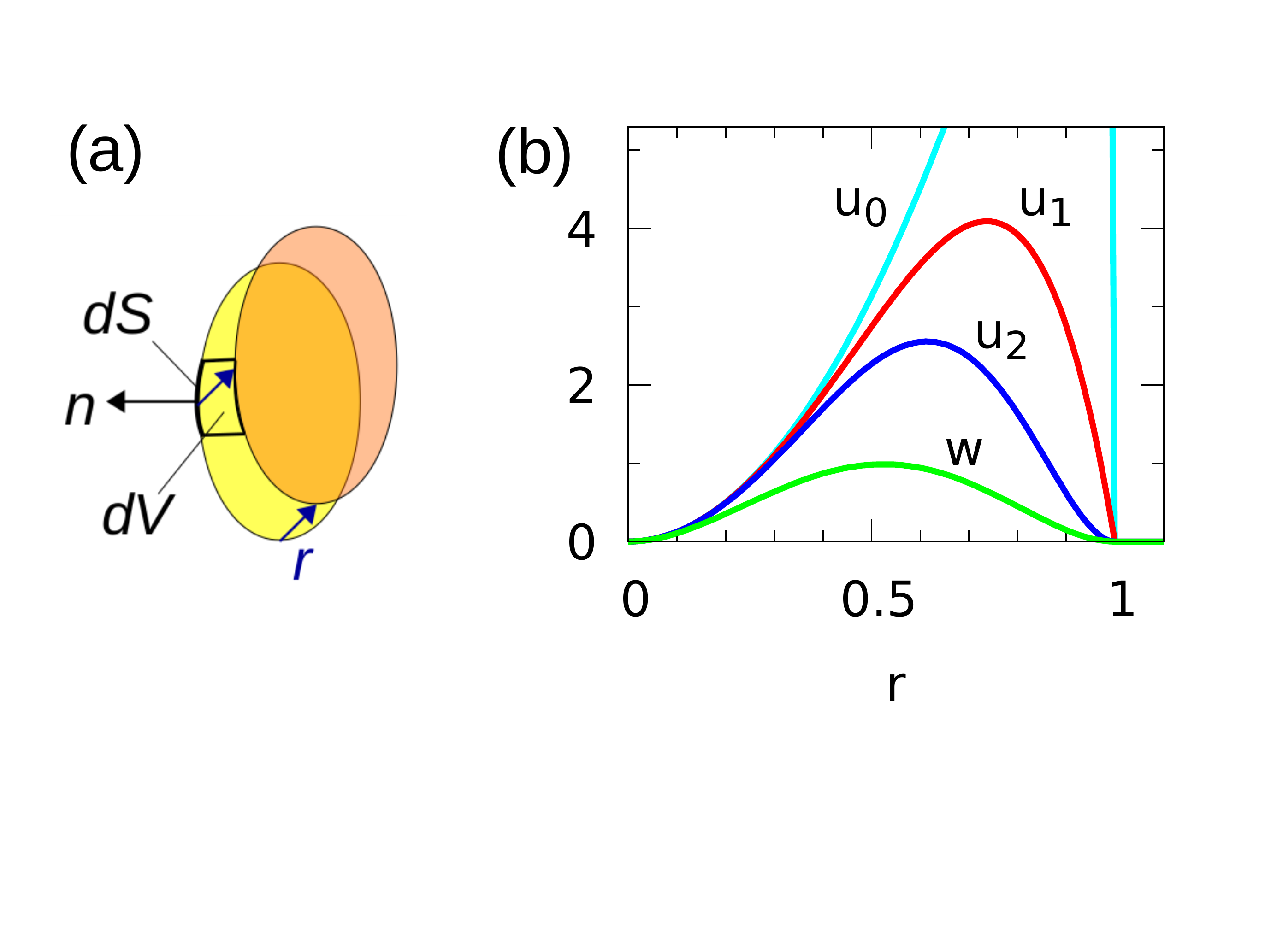}
\end{center}
\caption{
(a)  Illustration of overlap volume $\tau({\bf r})$ (orange)
  for a volume $V$ (yellow) and a small vector ${\bf r}$ (blue).
  At the surface element $dS$, $\tau$ is decreased from~$V$
  by $dV=-dS{\bf n}\cdot{\bf r}$, 
  where ${\bf n}$ is the surface normal.
(b)  Weight functions $u_k(r)$ and exact $w(r)$ for sphere, 
  plotted for $L=1$. See text for details.  }\label{w012}
\end{figure}

{\em Proof.}\/
We define
\begin{equation}\label{ur}
  y(r) = \frac{1}{4\pi V}\int d\Omega\; \tau({\bf r}) = \frac{w(r)}{4\pi r^2} \;.
\end{equation}
It is obvious from the definition
(Eq.~\ref{tau}) that $\tau(0)=V$ and so $y(0) = 1$.
We now determine the first order term of the $y(r)$ expansion around $r=0$.
Consider $\tau({\bf r})$ for a vector ${\bf r}$
small compared to the
linear dimension of the volume and to the curvature of the surface.
When $V$ is shifted by ${\bf r}$, the overlap volume $\tau$ decreases
at all surface points where ${\bf n}.{\bf r}<0$, where ${\bf n}$
denotes the (outward) surface normal.
Around such a point in a smooth surface region, the change in overlap
volume is $-dV=dS {\bf n}.{\bf r}$, where $dS$ is the surface element,
see Fig.~\ref{w012}~a.
So the total change is
$\Delta \tau({\bf r}) = \int' dS {\bf n}.{\bf r}$,
where the prime means that the surface integral is restricted to
points with ${\bf n}.{\bf r}<0$. Next, consider the opposite shift, $-{\bf r}$.
We get
$\Delta \tau(-{\bf r}) = -\int'' dS {\bf n}.{\bf r}$,
where the double prime denotes restriction to
points with ${\bf n}.{\bf r}>0$.
Summing the integrals due to both shifts, we obtain
\begin{equation}
\Delta \tau_\pm({\bf r})=
\Delta \tau({\bf r})+\Delta \tau(-{\bf r}) 
= -\int dS |{\bf n}.{\bf r}|   \;.
\end{equation}
Now the integral extends over the whole surface, because any point
has either ${\bf n}.{\bf r}<0$ or ${\bf n}.{\bf r}>0$ or ${\bf n}.{\bf r}=0$,
for which the contribution to the integral obviously vanishes.
We thus get
\begin{eqnarray}
\Delta y(r) &=& \frac{1}{4\pi V}\int d\Omega\; \Delta \tau({\bf r})
= \frac{1}{8\pi V}\int d\Omega\;\Delta\tau_\pm({\bf r})
\nonumber \\
& =& -\frac{1}{8\pi V} \int dS\int d\Omega\; |{\bf n}.{\bf r}| \;,
\end{eqnarray}
where the ${\bf n}$ corresponds to the surface ($dS$) integration,
and ${\bf r}$ is integrated over angles $\Omega$.
For fixed ${\bf n}$, we can easily perform the $\Omega$ integral:
\begin{equation}
\int d\Omega\; |{\bf n}.{\bf r}| = r \int d\Omega\; |\cos\theta|
= 2 \pi r \;,
\end{equation}
which is independent of ${\bf n}$. We get $\Delta y(r) = -{Ar}/({4V})$,
where $A=\int dS$ is the surface area.
This gives $y(r)$ to first order in~$r$:
\begin{equation}\label{uexpansion}
y(r) = y(0) + \Delta y(r) + {\cal O}(r^2)
= 1 - \frac{A}{4V}\;r + {\cal O}(r^2) \;.
\end{equation}
Together with the definition of $y(r)$, Eq.~(\ref{ur}), 
this completes the proof. It is valid for volumes of arbitrary shape,
as long as the surface is smooth almost everywhere,
in particular if the surface is piecewise smooth.

The function $y(r)=w(r)/(4\pi r^2)$
is dimensionless and thus invariant under a change of length scale
$r\rightarrow \alpha r$.
Therefore, $y$ depends on $r$ only in the combination $x=r/L$,
where $L\equiv 6V/A$ is the linear dimension of the system.
As a consequence, Eqs~(\ref{wtheorem},\ref{uexpansion})
are not only expansions in $r$, but also in $1/L$,
and the error is of order $1/L^2$.
This has important consequences for the
integrals~$G(L)$$\equiv$$G(V)$ in Eqs~(\ref{GV},\ref{Gw}),
in the limit $V,L\rightarrow\infty$.
From Eq.~(\ref{wtheorem})
we directly obtain the leading terms in a $1/L$ expansion
\begin{equation}\label{GL}
G(L) = \int_0^\infty h(r) w(r) dr = 
G^\infty + \frac{1}{L} F^\infty + {\cal O}(L^{-2})
\end{equation}
with $G^\infty$ given by Eq.~(\ref{Ginfty}) and
\begin{equation}\label{Finfty}
F^\infty = \int_0^\infty h(r)\left(-\frac{3}{2}r\right) 4\pi r^2 dr \;.
\end{equation}
In Eq.~(\ref{GL}), the shape only enters
through $L=6V/A$, i.e. the volume over surface ratio.
Expression~(\ref{Finfty}) is the universal 
surface term of $G(V)$ in Eq.~(\ref{GV}), valid for any shape
in the large volume limit.

The infinite volume KB integral, $G^\infty$ in Eq.~(\ref{Ginfty}),
provides the link to thermodynamic quantities,
but accurate computation of $G^\infty$ is difficult for two reasons.
First, the integral in Eq.~(\ref{Ginfty}) extends to $r\rightarrow\infty$
but in practice, the PCF is only known in a finite range $r<L$.
Second, when the PCF is obtained from molecular dynamics simulations of
{\it closed}\/ systems, it contains a systematic error even for $r<L$,
since the PCF is defined in the grand-canonical ensemble.
Here we focus only on the first point.
For the second problem, suitable corrections have been proposed
recently~\cite{bennaim,ganguly13,cortes16}.

We want to estimate $G^\infty$ in Eq.~(\ref{Ginfty}) from
the knowledge of $h(r)$ for $r<L$.
We look for an approximation (labeled by $k$) of the form
\begin{equation}\label{G0L}
G^\infty  \; \approx \;  {G_k}(L)=\int_0^L h(r) u_k(r)dr 
\end{equation}
with some weight function $u_k(r)$.
It is well known that the simple truncation of Eq.~(\ref{Ginfty}) at $r=L$
converges badly~\cite{nichols09,ganguly13}. This truncation
is denoted $G_0(L)$ with  $u_0(r)=4\pi r^2$~\cite{notation}.
In Ref.~\cite{kruger13}, we have proposed an extrapolation
$G_1(L)$, defined by $u_1(r)=4\pi r^2(1-x^3)$ with $x=r/L$,
which was obtained 
from a first order Taylor expansion in $1/L$
of the exact finite volume integral of the sphere
(Eqs~\ref{Gw},\ref{sphere}).
It was shown that the function $G_1(L)$ is more stable and 
converges faster than ${G_0}(L)$~\cite{kruger13,milzetti}.

Here, we derive a new extrapolation to $G^\infty$,
denoted $G_2(L)$, which converges much faster than
both~$G_0(L)$ and $G_1(L)$.
We rewrite Eq.~(\ref{GL}) as
\begin{equation}\label{Ginf1}
G^\infty \approx G(L) -\frac{1}{L} F^\infty \;.
\end{equation}
and calculate $G(L)$ exactly for a spherical volume using
$w(r)$ in Eq.~(\ref{sphere}).
The surface term $F^\infty$ must be approximated.
Defining ${\tilde h}(r)=-(3/2)rh(r)$, we can write
$F^\infty = \int_0^\infty {\tilde h}(r) 4\pi r^2 dr$, which has the same
form as $G^\infty$ in Eq.~(\ref{Ginfty}) except that $h$ is replaced
by ${\tilde h}$.
Therefore, we may use Eq.~(\ref{Ginf1}) again with
replacements $G\rightarrow F$, $h\rightarrow {\tilde h}$,
to obtain an approximation for $F^\infty$.
This yields
\begin{eqnarray}\label{Finf}
F^\infty&\approx&
\int_0^L {\tilde h}(r)w(r)dr-\frac{1}{L}
\int_0^\infty {\tilde h}(r)\left(-\frac{3}{2}r\right) 4\pi r^2dr \nonumber\\
        &\approx&
\int_0^L {\tilde h}(r)\left(1+\frac{3}{2}x\right)w(r)dr
\end{eqnarray}
where $x=r/L$ and in the last step, the infinite integral
has been replaced by the corresponding finite volume expression
for a sphere of diameter~$L$, with $w(r)$ given in Eq.~(\ref{sphere}).
Combining Eqs~(\ref{sphere},\ref{Ginf1},\ref{Finf}), we finally obtain
$G^\infty\approx G_2(L)$ with
\begin{eqnarray} \label{u2w}
  u_2(r) & =& w(r)  \left(1+\frac{3}{2}x+\frac{9}{4}x^2\right) \\ \label{u2}
       &=& 4 \pi r^2
\left( 1 - \frac{23}{8} x^3 + \frac{3}{4}x^4 + \frac{9}{8} x^5\right) \;.
\end{eqnarray}

In deriving Eq.~(\ref{u2w}) we have
applied Eq.~(\ref{Ginf1}) twice, in a recursive way.
It is easy to see that if we repeat the recursion $k$ times,
we obtain a weight function
$\tilde{u}_k(r) = w(r) \sum_{m=0}^k \left({3x}/{2}\right)^m $.
(We have $u_2\equiv {\tilde u}_2$ but we write
${\tilde u}_k$ in order to distinguish from $u_{0},u_{1}$ defined above.)
It is important to note that the series diverges
for $x>2/3$ and so the recursion
must be stopped at finite~$k$.
In numerical tests, we found best results for $k=2$.
Also note that because $w(r)$ contains a $x^3$ term,
the leading power of the correction $[\tilde{u}_k(r)/u_0(r)-1]$ is
$x^3$ for all $k\ge 2$ and consequently $G_k(L)-G^\infty\sim 1/L^3$. 
In other words, the leading $1/L^3$ behaviour of $G_2(L)$
cannot be improved by going beyond $k=2$.

We compare the convergence of the approximations $G_k(L)$
($k$=0,1,2) with a model PCF
of a liquid, given by~\cite{kruger13}
\begin{equation}\label{h}
  h(r) = 1.5\exp[(1-r)/\chi]\cos[2\pi(r-1.05)]/r
\end{equation}
for $r>0.95$ and $h(r)=-1$ for $r<0.95$.
The particle diameter ($\sigma$) has been chosen as the unit length.
\begin{figure}
\begin{center}
  \includegraphics[width=\columnwidth]{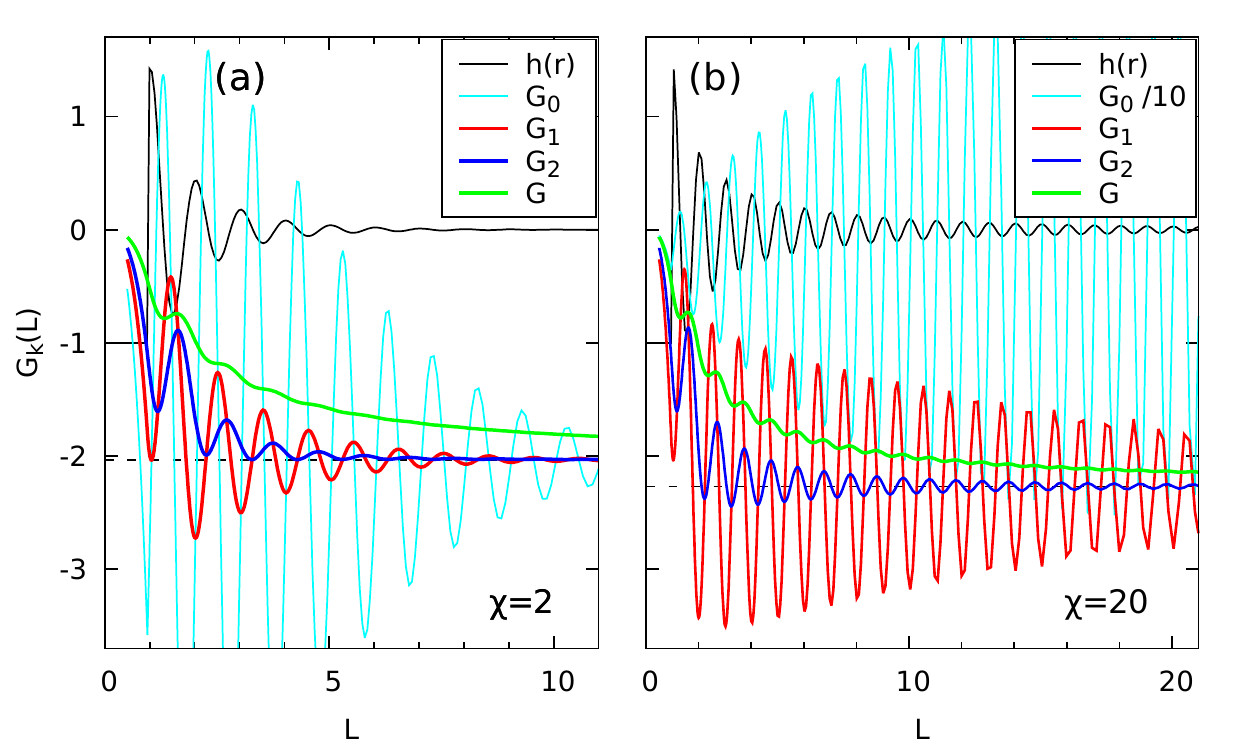}
\end{center}
\caption{
  Approximations $G_{0,1,2}(L)$
  to $G^\infty$ as a function of upper integration bound $L$
  for the PCF $h(r)$ of
  Eq.~(\protect\ref{h}) with $\chi=2$ (a) and $\chi=20$ (b).
  The finite volume integral $G(L)$ for a sphere of diameter~$L$ is also
  shown (green).
  }\label{gL2}
\end{figure}
The results for a short ($\chi=2$) and a
very large correlation length ($\chi=20$) are shown in Fig.~\ref{gL2}.
We see that $G_0$ has huge oscillations and thus poor convergence,
which is a serious problem in numerical
studies~\cite{nichols09,ganguly13,chiba16}.
The previous extrapolation $G_1$~\cite{kruger13}
oscillates much less than $G_0$ and converges faster.
For the new extrapolation $G_2$, the oscillations are reduced even more, 
resulting in a dramatically improved convergence.
For example, in the case $\chi=2$, 
for the error $|G(L)-G^\infty|/|G^\infty|$ to be smaller than 1\%,
we need $L>10$ with $G_1$ but only $L>7$ with $G_2$.
For $\chi=20$ the difference is even more striking:
$L>85$ is needed with $G_1$ but only $L>19$ with $G_2$.
The improvement of convergence by going from $G_0$ over $G_1$ to $G_2$
can be understood from the analytical properties of the weight functions
$u_k(r)$ shown in Fig.~\ref{w012}~b for $L=1$.
All functions coincide for small $r$ but they
have a very different behavior around $r=L$.
The function $u_0(r)$ is discontinuous at $r=L$, which explains why
the oscillations of $h(r)$ are hugely amplified in $G_0(L)$.
The functions $u_1(r)$ and $u_2(r)$ are continuous and so $G_1$ and $G_2$
fluctuate much less.
However, $u_1(r)$ has a discontinuous slope at $r=L$,
whereas $u_2(r)$ is smooth everywhere.
As a result, $G_2(L)$ has much weaker oscillations and thus
converges much faster than $G_1(L)$.

On more physical grounds, the poor convergence of $G_0(L)$
is due to the fact~\cite{kruger13} that it is not related to any
finite volume integral (Eq.~\ref{fluct}).
Since terms of order $A/V$ are absent in $u_0(r)=4\pi r^2$,
the corresponding finite volume would have zero surface
area, which is impossible.
Conversely, the good performance of $G_2(L)$ reflects the fact
that $G_2(L)$ is entirely based on properties of the
{\em exact}\/ finite volume integral $G(L)$~(Eqs~\ref{GV},\ref{sphere}).
Note that 
$G(L)$ (green line in Fig.~\ref{gL2}) is the smoothest of all $G$-functions,
but it approaches $G^\infty$ only slowly as $1/L$ (Eq.~\ref{GL}).
The new extrapolation $G_2(L)$ is also very
smooth, because $u_2(r)$ contains
the finite volume function $w(r)$ as a factor~(Eq.~\ref{u2w}).
However, $G_2(L)$ converges much faster than $G(L)$, namely as $1/L^3$,
as can be seen from the leading $x^3$ term in Eq.~(\ref{u2})
and as we have checked numerically.
As the computational costs of a molecular dynamics simulations scales at best
as $L^3$, it is clear that using the new extrapolation $G_2(L)$ will lead to a
large gain in computing time.

In summary, we have obtained the following general results on finite
volume KBIs and their convergence to the infinite volume limit.
(i) The reduction to a simple radial integral with analytic
weight function, which so far had been known only for spheres,
has been achieved for cubes and cuboids.
(ii) The leading term of the KBIs in a large volume expansion, 
only depends on the surface area-to-volume ratio.
This conjecture has been proved rigorously for volumes of
any shape.
(iii) A general expression for the surface term has been derived.
(iv) Different approximations for the infinite volume limit have been compared
and the reason why the running KBIs converge so poorly has been
elucidated.
(v) Based on the universal surface term,
a new extrapolation to the infinite volume limit has been derived,
which converges much faster than the approximations used so far.
The new extrapolation is expected to dramatically accelerate
numerical computation of KBIs.

We thank Jean-Marc Simon and Noura Dawass for many stimulating
discussions. TJHV acknowledges NWO-CW for a VICI grant.


\begin{thebibliography}{99}
\bibitem{notation}
Here we use a different notation than in
Ref.~\protect\cite{kruger13}. The correspondence is
$G_0(L)={\tilde G}^R$, $G_1(L)={\hat G}^R$ with $L=2R$.
\end{thebibliography}
\end{document}


\title{Size and shape dependence of finite volume Kirkwood-Buff integrals \\
 --- Supplemental Material --- }
\author{Peter Kr\"{u}ger}
 \affiliation{Graduate School of Engineering and
 Molecular Chirality Research Center, Chiba University,
1-33 Yayoi-cho, Inage,
 Chiba 263-8522, Japan}  
 {\small e-mail: {\tt pkruger@chiba-u.jp}}
\author{Thijs J. H. Vlugt}
\affiliation{
  Engineering Thermodynamics, Process \& Energy Department, Delft University of Technology,
  Leeghwaterstraat 39, 2628CB Delft, The Netherlands}

\date{\today}

\maketitle
\section*{Derivation of the function $T(r)$ of a cuboid}
\begin{figure} 
\begin{center}
\includegraphics[width=0.3\columnwidth]{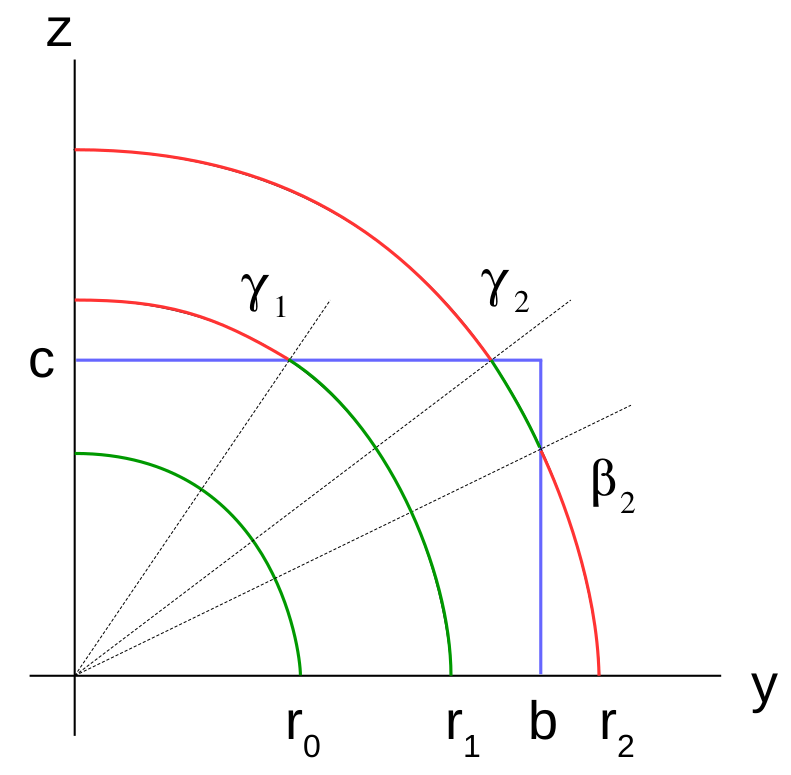}
\end{center}
\caption{Angular integration domains in the $yz$-plane.
  The green parts lie in the cuboid and are integrated over.
  The red parts lie outside of the cuboid and are subtracted.}
\label{fig1}
\end{figure}

Here, we calculate the function $T(r)$ for a cuboid of
side lengths $a\ge b\ge c$, i.e. we derive Eqs~(9,10) and the results in
Table~I of the main text.
We recall the definitions
\begin{equation}\label{tau}
  \tau({\bf r}) = \int_Vd{\bf r}_1\int_Vd{\bf r}_2 \delta({\bf r}_1 - {\bf r}_2 - {\bf r})
  \quad {\rm and} \quad  T(r)= \int d\Omega\; \tau(r,\Omega)   \;.
\end{equation}
We choose coordinate axes $x$, $y$, $z$ along $a$, $b$, $c$.
The calculation of $\tau({\bf r})$ is trivial. We have
\begin{eqnarray}\label{tauxyz}
  \tau(x,y,z) = (a-|x|)(b-|y|)(c-|z|)
  \Theta(a-|x|)\Theta(b-|y|)\Theta(c-|z|) \;,
\end{eqnarray}
where $\Theta(x)$ is the Heaviside step function.
For the integral over the solid angle~$\Omega$ in Eq.~(\ref{tau}), we use
spherical coordinates,
$x=r\sin\theta\cos\phi$, $y=r\sin\theta\sin\phi$, $z=r\cos\theta$.
As $\tau$ is an even function of $x$, $y$, $z$, it is enough to
consider $x,y,z>0$ and so,
\begin{equation}\label{Tr}
T(r) = 8\int_0^{\pi/2} d\theta\sin\theta \int_0^{\pi/2}d\phi
\;\eta(r,\theta,\phi)\vartheta(r,\theta,\phi)
\end{equation}
where
\begin{eqnarray}\label{}
 \eta(r,\theta,\phi) &=&
 (a-r\sin\theta\cos\phi)(b-r\sin\theta\sin\phi)(c-r\cos\theta)
 \\
  \vartheta(r,\theta,\phi) &=&
  \Theta(a-r\sin\theta\cos\phi)\Theta(b-r\sin\theta\sin\phi)\Theta(c-r\cos\theta)\;.
\end{eqnarray}
We consider the integral
\begin{equation}\label{}
U(r,\theta) = 8\int_0^{\theta} d\theta' \sin\theta'
\int_0^{\pi/2}d\phi \;\eta(r,\theta',\phi) \;.
\end{equation}
The $\phi$-integral is trivial and yields 
\begin{equation} \label{Urt}
  U(r,\theta) = 8\int_0^{\theta} d\theta' \sin\theta' X(\theta')
  \end{equation}
  where
\[
X(\theta)=
\frac{\pi}{2} abc -r\left(\frac{\pi}{2} ab\cos\theta+ac\sin\theta+bc\sin\theta
\right)
+
r^2 \left(
a\cos\theta\sin\theta+b\cos\theta\sin\theta+\frac{c}{2}\sin^2\theta
\right)
- \frac{1}{2}r^3 \cos\theta\sin^2\theta \;.
\]
\begin{table}
\begin{tabular}{c|c|c|c|c|c|c}
  $f(\theta)$
  &  1 & $K$ & $S$    &  $KS$   & $S^2$    & $KS^2$  \\ \hline
  $\int_0^{\theta}d\theta'\sin\theta'f(\theta')$
  & $1-K$ & $\frac{1}{2}S^2$ & $\frac{1}{2}(\theta-KS)$ 
  & $\frac{1}{3}S^3$     &     $\frac{1}{3}(K^3-3K+2)$     
  &   $\frac{1}{4}S^4$\\
  \end{tabular}
\caption{Elemental integrals over $\theta'$ appearing in Eq.~(\ref{Urt}).
Here $K\equiv\cos\theta$, $S\equiv\sin\theta$.}\label{tab1}
\end{table}
The $\theta'$ integration in Eq.~(\ref{Urt}) is also straightforward,
see Table~\ref{tab1}. We find
\begin{eqnarray}
U(r,\theta)
&=& {4\pi}abc(1-\cos\theta)
- {2\pi}abr\sin^2\theta
- 4(a+b)cr(\theta-\cos\theta\sin\theta)
\nonumber \\
&+& \frac{8}{3}(a+b)r^2\sin^3\theta
+ \frac{4}{3}cr^2(\cos^3\theta-3\cos\theta+2)
- r^3\sin^4\theta
\end{eqnarray}\label{Urtheta}
We assume $a \ge b \ge c$.
For $r<c$ (such as $r_0$ in Fig.~\ref{fig1}), the integration sphere
lies entirely in the cuboid. Then $\vartheta(r,\theta,\phi)=1$ for all
$\theta$, $\phi$ and we get
\begin{equation}\label{i0}
  T(r) = U(r,\pi/2) \equiv P(r)\;, \qquad {\rm for\ } r<c
\end{equation}
where 
\begin{equation}\label{P}
P(r)
= {4\pi}abc
- {2\pi}(ab+ac+bc)r
+ \frac{8}{3}(a+b+c)r^2
- r^3
\end{equation}
When $r>c$, some parts of the sphere exceed the cuboid
and so the integration domains over $\theta$ and $\phi$ become restricted.
First, consider $c<r<b$, such as $r_1$ in Fig.~\ref{fig1}.
The part to be subtracted from the sphere corresponds to two
spherical caps around $\pm z$ with half angle $\gamma_1=\arccos(c/r_1)$
(red line in Fig.~\ref{fig1}.) 
The corresponding integral is $U(r_1,\gamma_1)$ and so we obtain
\begin{equation}
  T(r) = U(r,\pi/2) - U(r,\arccos(c/r))\equiv P(r)-Q(r;a,b,c)\;, \qquad {\rm for\ } c<r<b
\end{equation}
where
\begin{eqnarray}\label{Q}
  Q(r;a,b,c) & =&
  4\pi abc
- \left( {2\pi}ab + {4\arccos(c/r)}(a+b)c + 2c^2\right)r
+ \frac{8}{3}cr^2 -r^3
\nonumber
\\ 
& +& \left(\frac{1}{3}c^4-{2\pi}abc^2\right) \frac{1}{r}
+\frac{4}{3}(a+b)(c^2+2r^2) \sqrt{1-\frac{c^2}{r^2}} 
\end{eqnarray}
The dependence of $Q$ on $a$,$b$,$c$ (in this order) is made explicit,
because below we shall use $Q(r;c,a,b)$ which denotes Eq.~(\ref{Q})
with $a$,$b$,$c$ permuted.
Second, consider $b< r< \sqrt{b^2+c^2}$, such as $r_2$ in Fig.~\ref{fig1}.
In this case, the sphere exceeds the cuboid also around the $y$-axis.
The new parts to be subtracted are spherical caps of
half-angle $\beta=\arccos(b/r)$. The corresponding
$\theta$-range is $\pi/2-\beta<\theta<\pi/2$,
but the $\phi$-integration is also restricted.
The integral
can be performed more easily by rotating the coordinate system 
such that the new axes are $z'=y$, $x'=z$, $y'=x$, $z'=y$,
i.e. $b$ lies along $z'$.
For this ``b-cap'' integration we then obtain
the same expression as $U(r,\arccos(c/r))$ above
except for the replacements $c\rightarrow b\rightarrow a$
(implying $\gamma\rightarrow\beta=\arccos(b/r)$)
due to the permutation of the axes. Hence
\begin{equation}
  T(r) = P(r) - Q(r;a,b,c) - Q(r;c,a,b)
  \qquad {\rm for\ } b<r<{\rm min}\{a,\sqrt{b^2+c^2}\}
\end{equation}
For $a<r<\sqrt{b^2+c^2}$, two more caps exceed the cuboid 
around the $x$-axis
(not shown in the $yz$-plane of Fig.~\ref{fig1}.)
The corresponding integral is again obtained from Eq.~(\ref{Q})
upon replacements $c\rightarrow a\rightarrow b$, so
 \begin{equation}\label{i0cba}
   T(r) = P(r) - Q(r;a,b,c) - Q(r;c,a,b)- Q(r;b,c,a)
  \quad {\rm for\ } a<r<\sqrt{b^2+c^2}
 \end{equation}
The results are summarized in Table~I of the main text.